\documentclass[twocolumn,prb,showpacs]{revtex4}
\usepackage{psfrag}
\usepackage{graphicx}
\usepackage{amsmath}
\usepackage{amssymb}

\usepackage{color}
\begin{document}
\title{Electron transfer through a single barrier inside a molecule: from strong to weak coupling}
\author{Robert Stadler$^1$, J\'{e}r\^{o}me Cornil$^2$ and Victor Geskin$^2$} 
\affiliation{$^{1}$Department of Physical Chemistry, University of Vienna, Sensengasse 8/7, A-1090 Vienna, Austria\\
$^{2}$Laboratory for Chemistry of Novel Materials, University of Mons, Place du Parc 20, B-7000 Mons, Belgium}

\date{\today}

\begin{abstract}
In all theoretical treatments of electron transport through single molecules between two metal electrodes, a clear distinction has to be made between a coherent transport regime with a strong coupling throughout the junction and a Coulomb blockade regime in which the molecule is only weakly coupled to both leads. The former case where the tunnelling barrier is considered to be delocalized across the system can be well described with common mean-field techniques based on density functional theory (DFT), while the latter case with its two distinct barriers localized at the interfaces usually requires a multideterminant description. There is a third scenario with just one barrier localized inside the molecule which we investigate here using a variety of quantum-chemical methods by studying partial charge shifts in biphenyl radical ions induced by an electric field at different angles to modulate the coupling and thereby the barrier within the $\pi$-system. We find steps rounded off at the edges in the charge versus field curves for weak and intermediate coupling, whose accurate description requires a correct treatment of both exchange and dynamical correlation effects is essential. We establish that DFT standard functionals fail to reproduce this feature, while a long range corrected hybrid functional fares much better, which makes it a reasonable choice for a proper DFT-based transport description of such single barrier systems. 
\end{abstract}
\pacs{73.63.Rt, 73.20.Hb, 73.40.Gk}
\maketitle

\begin{section}{Introduction}\label{sec:intro}

An important aspect in the vibrant new field of single-molecule electronics is the classification of the tunnelling process with respect to the coupling between the molecule and metallic leads. Coherent tunnelling (CT) with an electron passing through the junction in one step (strong coupling) is contrasted with Coulomb blockade (CB) with the electron resting on the molecule in a two-step process (weak coupling), which can be distinguished experimentally~\cite{kubatkin}-~\cite{kouwen}. The theoretical description of electron transport through single-molecule junctions commonly involves the combination of a non-equilibrium Green's function (NEGF) formalism~\cite{keldysh} for defining the semi-infinite boundary conditions of the transport problem and density functional theory (DFT) with typically a semi-local exchange-correlation (XC) functional for a reasonably accurate description of the ground-state electronic structure of the junction at an atomistic level.~\cite{atk}-~\cite{kristian} While this approach is known to reproduce the smooth current-voltage (I/V) behaviour characteristic of CT, it does not capture the distinct steps in I/V curves which are the signature of CB in experiments. There are two different schools of thought for explaining this deficiency. The first one relates it to the existence of spurious self-interaction errors (SIE) in DFT calculations with semi-local functionals~\cite{sanvito1,wang,yang}, i.e. essentialy to the approximation for exchange in the XC functional, while the other blames the single determinant nature of DFT and therefore its inadequacy in describing correlation~\cite{datta}-~\cite{ghosh,multidet}.

\begin{figure}
    \includegraphics[width=0.95\linewidth,angle=0]{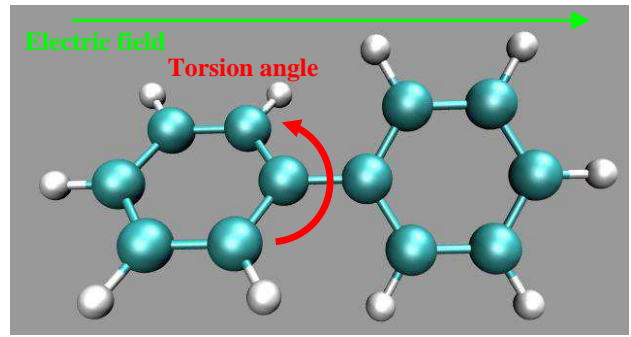}
      \caption[cap.Pt6]{\label{fig.biphenyl}In a biphenyl molecule, the coupling between the $\pi$ systems on its two phenyl halves is defined by the torsion angle between the rings: the more coplanar the stronger the coupling. We charge this molecule by subtracting a single electron and apply an external electric field for shifting partial charges between the two components of the resulting cation.}
    \end{figure}

In a way, the distinction between CT and CB made only from the coupling to the electrodes considers the molecule at the heart of the junction as a featureless quantum dot. For molecular electronics, however, it is particularly relevant to pay attention to the contribution of the molecule to the transmission of the junction and its theoretical description. It is not unreasonable to consider that similar problems to those encountered in DFT treatments of the double barrier situation might also occur for electron transport through a single barrier. Therefore, we focus in this work on the description of a single intramolecular barrier, by studying a molecular model system and performing Hartree-Fock (HF), post-HF and DFT based quantum-chemical calculations to describe the charge shifts induced by an external electric field. In particular, we investigate the charge shifts inside biphenyl radical ions, with the torsion angle between the two phenyl rings acting as a parameter for the coupling of the two $\pi$-conjugated rings (see Fig.~\ref{fig.biphenyl}). 

For the interpretation of our results we establish a link between our calculations and an alternative perspective on the limitations of DFT presented by Cohen and co-workers, recently~\cite{cohen1}-~\cite{cohen4}. In this framework DFT is considered to be exact and universally applicable in principle but the XC functionals currently in use are argued to suffer from two fundamental shortcomings, namely a delocalization error and a static correlation error~\cite{cohen1}-~\cite{cohen4}. The former is closely linked to SIE and completely absent in HF, while the latter affects HF even more than conventional DFT. For both problems, chemically simple but methodologically still challenging test systems have been identified, namely the dissociation curves of H$_2^+$ and H$_2$, respectively. In this picture, it follows naturally that for H$_2^+$ only exchange matters because the system contains only a single electron, while for H$_2$ at inter-atomic distances far beyond the optimal bonding length, the non-dynamic (or strong) correlation of the two unpaired electrons on the separated H atoms has to be correctly described. 

While Cohen et al.~\cite{cohen1}-~\cite{cohen4} and others~\cite{daul,johnson} developed their ideas for the derivation of better XC functionals for general purposes, they also stated that their systematics should be relevant in particular for the description of electron transport problems. This suggestion has not been followed up by transport specific calculations so far while such studies appear to be necessary because: i) as outlined above, in electron transport theory, there is still no agreement within the field about the relative importance of exchange or correlation, respectively, and ii) there is still no NEGF-DFT approach available which can reproduce steps in I/V curves correctly in general. While methods exist, which go beyond a mean-field description~\cite{mayou,kristian1,hybertsen,greer}, these techniques are computationally demanding and do not lend themselves easily to chemical interpretations in terms of molecular orbitals (MOs). It is therefore highly desirable to find out if there is an available XC functional which is able to describe steps in I/V curves correctly within NEGF-DFT. In a recent study of the Anderson junction this question is addressed using a Bethe ansatz~\cite{stafford} but the findings are limited to a single level occupied with a single electron and need to be complemented by highly accurate quantum chemical calculations on real molecular systems. We consider our current work to be a move in this direction, by showing that common density functionals fail to reproduce the expected intra-molecular charge shifts correctly at weak coupling in particular, while a long-range corrected functional behaves reasonably. 

The paper is organized as follows: in the next two sections, we establish a direct link to the work of Cohen et al.~\cite{cohen1}-~\cite{cohen4} by arguing that the artefacts in the description of charge shifts are related to the H$_2$ and H$_2^+$ dissociation problems for the biphenyl neutral molecule and cation, respectively. In Section~\ref{sec:xc} we present DFT calculations on charge shifts in biphenyl cations (we obtained quite similar results for anions which we do not discuss explicitly in this article) using a variety of common XC functionals with the near-coupled cluster quadratic singles and doubles method (QCISD) as reference. We find that long-range corrected (LC) hybrid functionals~\cite{scuseria1}-~\cite{scuseria3} provide satifactory solutions for the whole coupling range. Finally, we provide a summary where we also discuss the relevance of our results for NEGF-DFT-based electron transport calculations. 

\end{section}

\begin{section}{Shifting charges inside molecular cations and neutral molecules as a probe for coupling in model systems}\label{sec:cohen}

In order to establish a link between the dissociation of molecules and ions on the one hand and electron transport through single molecule junctions on the other hand, we take an intermediate step and pose an associated question, namely how dissociation can be linked to shifting charges inside single molecules with an electric field. First we discuss how such charge shifts should evolve in molecules or ions composed of two identical halves, such as in H$_2$ and biphenyl (BP). If no field is applied, the MOs contain equal contributions from atomic orbitals (AOs or, in general, fragment orbitals which we will also call AOs for the sake of simplicity in the following) of both parts by symmetry, and hence the charge is equally distributed regardless of the strength of the coupling. If a longitudinal electric field (EF) is introduced the left and the right sides experience different potentials,  the AO levels are only shifted upwards in energy on one side and downwards on the other side, which will influence the composition of the MOs accordingly (in a first approximation neglecting the polarization of AOs). 

When the coupling between the parts is sufficiently strong, the impact of the applied field will be an unequal contribution of the left and right AOs to all MOs, that is a polarization of the system, so that the composition of the MOs will change smoothly with EF. This gradual polarization of occupied MOs, in particular of the HOMO, induces a continuous shift of fractions of an electron from one side to the other. 
The weak coupling limit is more challenging for the description of charge shifts inside molecules. For zero coupling between the parts, in the absence of EF, all the MOs of our two-component system come in degenerate pairs arising from a symmetric and antisymmetric linear combination of the left and right AOs of the same type and also equal in energy to the respective AOs on both parts, since hybridization effects are ruled out by definition. If the symmetry constraints are lifted by applying EF, the MOs in every pair are split and take the shape of their constituent AOs. In the weak coupling limit there are fundamental differences between systems with even and odd numbers of electrons, which we discuss below with the help of Figs.~\ref{fig.h2_scheme} and~\ref{fig.biph_scheme}. 

For the neutral H$_2$ and BP molecules, which contain an even number of electrons, the process of shifting charges involves mainly the HOMO and the LUMO which in the zero coupling limit are reduced to the corresponding AOs on each half. Depending on the intensity of the applied field, there are two possible ground states obtained either by populating each orbital with one electron, or by populating the lower one with both electrons. For H$_2$ (Fig.~\ref{fig.h2_scheme}), the open-shell configuration with a field applied (panel E) has the same level occupation as the zero-field ground-state (panel B) with one electron on each side until the potential difference $\Delta$V (induced by EF and increasing with it) between the AO levels  exceeds the charging energy $\Delta$U. At this point the system is transformed to a closed shell configuration with both electrons on one side (panel D), where an electron "jumps" from the higher lying singly occupied level to the lower lying one, making the latter doubly occupied. 

We use a model system with four electrons in the frontier orbitals to mimic the behaviour of the $\pi$ system of a neutral BP molecule with the phenyl rings perpendicular to each other as indicated in panels B, D and E of Fig.~\ref{fig.biph_scheme}, where only the local HOMOs and LUMOs of the phenyls are taken into account). The zero-field ground state (panel B) is closed-shell, and EF (shifting all the occupied and vacant levels at each side in the same direction) has to reduce the gap ($\Delta_{gap}$) between the right HOMO and the left LUMO (panel E) until an electron transfer between these two levels creating two SOMOs at the opposite sides (panel D) becomes favorable. In this case, the electron jump leads to a transition from a closed to an open shell configuration. We note that the relation between $\Delta$U and $\Delta_{gap}$ with respect to $\Delta$V depends on the theoretical framework in this case, since the charging energy is contained in the energy gap for HF but contrarily it needs to be explicitly added to the Kohn Sham gap for arriving at addition energies within DFT~\cite{first,second}. For our purpose in this article, this particular distinction is not relevant since we focus on cations. In order to summarize the common features between the two cases with an even number of electrons, we conclude that for the electron to jump from one side to the other a charging energy $\Delta$U has to be overcome by the applied field EF.

\begin{figure}
    \includegraphics[width=0.95\linewidth,angle=0]{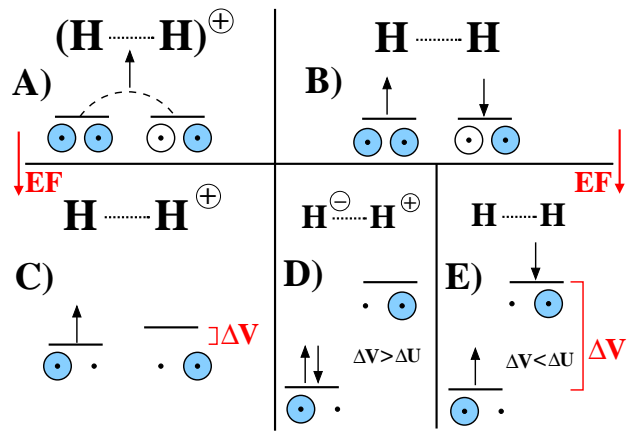}
      \caption[cap.Pt6]{\label{fig.h2_scheme}Schematic orbital diagram of the shift of electrons in H$_2^+$ (panels A and C) and H$_2$ (panels B, D, and E) for the limit of zero coupling between the orbitals before (A, B) and after (C, D and E) the application of a longitudinal electric field. For H$_2$ a critical voltage $\Delta$V exceeding the charging energy $\Delta$U has to be applied (D), while an infinitesimally small $\Delta$V is enough for H$_2^+$ (C). The corresponding MO patterns are shown below their energy levels, where positive and negative signs of the wavefunction are illustrated as blue and white spheres, respectively, and the position of the nuclei marked by black dots.}
\end{figure}

  \begin{figure}   
  \includegraphics[width=0.95\linewidth,angle=0]{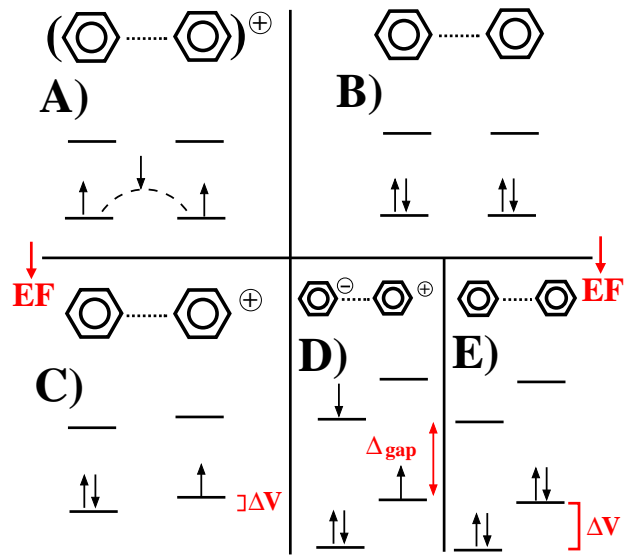}   
  \caption[cap.Pt6]{\label{fig.biph_scheme}Schematic orbital diagram of the shift of electrons in a longitudinal electric field between the $\pi$ systems of the two phenyl rings in a neutral biphenyl molecule (panels B, D, and E) and cation (panels A and C) for the limit of zero coupling. The portrayed orbitals are the local HOMOs and LUMOs on both rings, respectively. While for the neutral molecule the threshold $\Delta$V has to be exceeded (incorporating both the HOMO-LUMO gap and the charging energy) for a full charge transfer (D), an infinitesimally small electric field is enough for the cation (C).}
  \end{figure}     

For the cations H$_2^+$ and BP$^+$ with one and three electrons in the model system, respectively, for zero coupling and zero field, a single electron occupies one MO of the degenerate pair (see the upper left side panels of Figs.~\ref{fig.h2_scheme} and~\ref{fig.biph_scheme}). Due to symmetry, the probability to find the electron on either of the two AOs (halves of the cation) is equal. By applying EF this symmetry is lifted along with the degeneracy of the AOs and the electron becomes localized on one side. The spatial charge distribution then crucially depends on how localized or delocalized the fragment orbitals are. It is essential for the arguments in our paper that for the breaking of symmetry caused by the physical boundary conditions imposed by the external field an infinitesimally small $\Delta$V is sufficient and no charging energy $\Delta$U has to be provided or in other words in the zero coupling limit no resonance energy is lost by localization in the more stable of two quasi-degenerate structures. This distinguishes the systems with an odd number of electrons from those with an even number significantly.

In order to elucidate the connection between the two different physical phenomena of shifting charges inside molecules or ions and their dissociation by stretching bonds, it is worth noting that there are two key parameters in both cases determining the impact of their electronic structure: (i) the relative position of the AO levels of the isolated fragments and (ii) the coupling between their AOs. In the course of dissociation, for a chosen molecule, (i) remains constant being determined by the chemical nature of the fragments, while (ii) is continuously varied being a function of the distance. The parameter landscape of (i) can also be scanned in the case of dissociation, although discontinuously, by comparing the curves of various molecules. In contrast, when EF is applied to a given molecular system, (ii) remains constant (in a first approximation, neglecting the polarizing effect of the field on the shape of AOs), while (i) is varied continuously. If this is done for various couplings, it means that the scope of (ii) is also probed. Therefore, the entire parameter space explored in the description of the electronic structure of dissociating molecules is equivalent to that when shifting charges inside molecules with a longitudinal electric field and a variable torsion angle to modulate the electronic coupling. It is therefore reasonable to assume that methodological findings on the dissociation of H$_2^+$ and H$_2$ are also relevant for the applicability of the investigated methods to the problem of charge shifting with an external field. This conjecture, however, has to the best knowledge of ours never been explicitly tested which is a key motivation for the work presented in this article.
\end{section}
  
\begin{section}{From model systems towards DFT calculations}\label{sec:dft}

When performing DFT calculations of the dissociation potential energy curves of H$_2$ and H$_2^+$, respectively, two distinctly different scenarios can be identified with regard to the total energy for a large separation of the two H atoms~\cite{cohen2}. 

For H$_2^+$, DFT finds the state with the single electron in the system delocalized on the two H positions always overstabilized (delocalization error) vs. the state in which it is localized at one of the hydrogens, while in the low coupling (dissociation) limit these situations should tend to degeneracy. This deficiency results in dissociation to a state with fractional charges (which in itself is not an error for H$_2^+$, but can lead to errors for other diatomics~\cite{scuseria4}) and a binding energy that is far too low~\cite{daul}. It has been shown that for obtaining the correct physical relation between the total energies of atoms with integer and fractional charges, the XC-potential would need to exhibit a derivative discontinuity, whereas the common semi-local functionals do not~\cite{perdew1,perdew,chan}. Since this problem is related to SIE, HF calculations do not suffer from it (the self-interaction contributions of the Hartree and exchange potentials cancel out exactly), and therefore the binding energy of H$_2^+$ can be obtained exactly with HF. 

For H$_2$ the situation is very different. Here, the problem for mean-field computations is due to an insufficient description of the correlation between the spins of the two electrons~\cite{cohen2} in both DFT and HF (static correlation error). The latter case which corresponds to shifting charges in a neutral molecule (see the right side panels of Figs.~\ref{fig.h2_scheme} and~\ref{fig.biph_scheme}) closely resembles the electron jumps between molecules which we described in Ref.~\cite{multidet}. 
  
As our aim is to model the electron transport through a single intra-molecular barrier, we argue that the cations on the left sides of Figs.~\ref{fig.h2_scheme} and~\ref{fig.biph_scheme} are closer to such a setup than the neutral molecules, because an electron moving through a junction is provided by the electrodes. The injection of a carrier into the molecule can involve a charging energy but only if the molecule is weakly coupled to the electrodes. This is not the case we want to investigate in our present article, where we are only concerned with the effect of a single intra-molecular barrier and therefore assume that the electron (or in our study the hole, which is equivalent) has already reached the molecule and hence focus on cations. 

For H$_2^+$ and BP$^+$, which are open-shell systems as a whole, with and without the electric field, a closed-shell description becomes formally inapplicable. However, while the correlation effects are strictly inexistent for H$_2^+$, which allows unrestricted HF (UHF) to describe it properly~\cite{daul,bally}, the situation can be more complicated for BP$^+$. Although static correlation which is the main source of error for H$_2$ (and neutral BP) should be absent for the biphenyl cation as well as for H$_2^+$, BP$^+$ is not a one-electron system, so that dynamical correlation can become significant due to the interaction of the electron in the singly occupied orbital with all other electrons in the radical ion. Such effects have already been identified in the simpler but related case of allyl radicals~\cite{aquilante} and have also been discussed from the perspective of artificial symmetry breaking or the competition between spin polarisation and spin delocalization~\cite{glaser}.

The question now is how DFT with different XC-functionals deals with this situation. It is not easy to predict a priori and addressing this issue is the main objective of this article. Because the delocalization error is known to artificially stabilize the delocalized system for dissociation, leading e.g. to spurious fractional charges on the fragments~\cite{scuseria4}, it is natural to assume that for EF induced charge shifts, the same delocalization error makes the polarisation of a molecular cation more gradual (or in other words the charge distribution between its halves more equal), since in both cases the deficiency is due to an overestimation of the coupling throughout the molecule. 

For the first-principles calculations presented in the following, H$_2^+$ is an impractical system because the coupling between the atoms, which is the crucial quantity to probe the connectivity of a frontier MO relevant for electron transport, can only be varied by their distance. For BP$^+$ on the other hand, the coupling between the local $\pi$ electrons can be controlled by varying the torsion angle between the rings, which can also be adjusted by chemical means thereby allowing for its use as an experimentally trackable coupling parameter for electron transport measurements in single molecule junctions~\cite{hybertsen1,mayor}.

\end{section}

\begin{section}{Finding a suitable XC-functional for shifting charges in a biphenyl cation}\label{sec:xc}

\begin{figure*}   
\includegraphics[width=0.95\linewidth,angle=0]{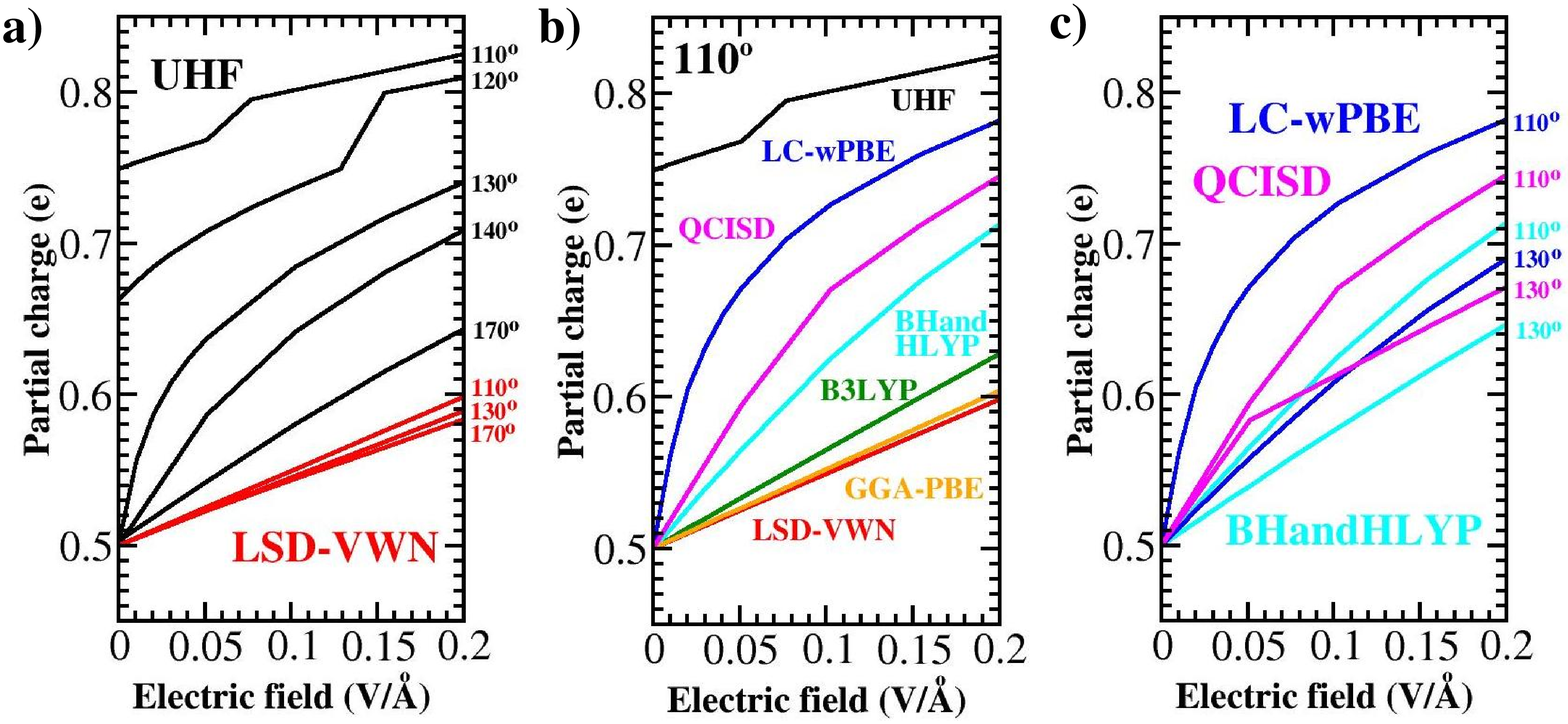}   
\caption[cap.Pt6]{\label{fig.charge}Partial positive charges on one of the two phenyl rings in a biphenyl cation induced by an electric field as calculated: a) for various torsion angles between the rings with UHF (black lines) and DFT making the LSD approximation in a VWN parameterisation (red lines); b) at a torsion angle of 110$^{\circ}$ with UHF (black), QCISD (magenta) and DFT with a variety of different XC-functionals, namely LSD-VWN (red), GGA-PBE (orange), B3LYP (green), BHandHLYP (cyan) and LC-$\omega$PBE (blue); c) for torsion angles 110$^{\circ}$ and 130$^{\circ}$ with QCISD (magenta) and DFT using LC-$\omega$PBE (blue) and BHandHLYP (cyan) as parameterisation of the XC-functional.
}
\end{figure*}     

All numerical results presented in this section are based on first-principles calculations performed with the GAUSSIAN09 package~\cite{gaussian}. We use this code because it allows for a direct comparison (i.e. within the same computational setup and using the same basis sets) of DFT based techniques with semi-local or hybrid XC-functionals with HF and post-HF implementations of wavefunction theory. In Fig.~\ref{fig.charge}a, we plot charge vs. voltage curves calculated with UHF (black lines) for a biphenyl cation for a variety of torsion angles ranging from 110$^{\circ}$ to 170$^{\circ}$ in the optimized geometry of a neutral BP molecule with the respective torsion imposed as a constraint, i.e., from very weak to very strong coupling between the $\pi$ systems on the two phenyl rings, respectively, as obtained from a summation of Mulliken charges. We note that for weak coupling (110$^{\circ}$-120$^{\circ}$) not only the smallest field we apply already breaks the symmetry between the two components and the charge localizes on one of the two phenyl rings but the zero-field charge distribution also has a broken symmetry, because the symmetric UHF solution turns out to be unstable, i.e. higher in total energy than the asymmetric one. The singly filled MO (SOMO) in particular always tends to localize preferentially on one of the two rings. For strong coupling (170$^{\circ}$), this SOMO is well delocalized over both components also for high fields and charge is therefore transferred continuously by a polarisation of this orbital. As can be seen in Fig.~\ref{fig.charge}a, this polarizability gradually increases as the driving force delocalization decreases for higher coupling/torsion angle.

Although UHF describes exchange interactions exactly, its symmetry breaking problem at zero field makes it a questionable bechmark even for calculating partial charges at a finite voltage where the field is breaking the symmetry as a physical boundary condition anyway. Though counterintuitive, this is due to the continuity of the finite field solution with the artificially asymmetric zero field solution. This implies that if the zero field solution is wrong, the finite field solution has to be wrong as well. At zero field and with the positions of the nuclei fixed the geometry of the wavefunction has to be symmetric because of the symmetry of the molecule, so that we need to find a method providing such a solution as a benchmark. Sometimes this can be achieved by increasing the size of the basis set but we checked that it is not the case here. Since the complete active space self-consistent field (CASSCF) and configuration interaction singles and doubles (CISD) methods still break electronic symmetry without bias to a significant degree, we chose QCISD calculations as a benchmark, a method for which it has been demonstrated for allyl radicals that higher order correlation effects are well described~\cite{glaser}. We checked that for our system the symmetry breaking tendency of this method is negligible and found that its main effect on the charging curves compared to UHF at an angle of 110$^{\circ}$ is to round off the edges of the steps considerably.

DFT calculations within the local spin density approximation (LSD) and using the Vosko-Wilk-Nusair (VWN) parameterisation~\cite{vwn} for the XC-funtional do not suffer from the symmetry breaking problem but its deviation from the physically correct curves are equally serious. Within this framework the partial charge curves become linear regardless of the torsion angle (red lines in Fig.~\ref{fig.charge}a), which is due to the lack of a derivative discontinuity in the XC-potential, and leads to an overstabilisation of fractional charges on both phenyl rings compared to a distribution of integer charges~\cite{perdew1,cohen2}. This deficiency can also be understood in terms of a SIE-induced delocalization error~\cite{cremer,siegbahn}, even for a torsion angle of 110$^{\circ}$, where the SOMO is almost equally ditributed on both rings at high fields with LSD-VWN, since delocalized states are always lowered in energy artificially. 

In Fig.~\ref{fig.charge}b we compare charging curves for an individual phenyl ring within a biphenyl cation at a torsion angle of 110$^{\circ}$, as calculated with a variety of different XC-functionals and measure them against QCISD as a physically correct benchmark. It can be seen that the generalized gradient approximation (GGA) with a Perdew-Burke-Ernzerhof (PBE)~\cite{pbe} parameterisation for the XC-functional as well as the hybrid exchange approach B3LYP~\cite{b3lyp1,b3lyp2} provide only incremental quantitative improvements when compared with LSD-VWN. BHandLYP, with 50\% of HF exchange, approaches the QCISD curve somewhat closer, still underestimating significantly the polarization induced by the field. We find, however, a quite different behaviour with the long-range corrected hybrid functional LC-$\omega$PBE~\cite{scuseria1}-~\cite{scuseria3} which, however, somewhat overshoots with respect to the QCISD polarisation at 110$^{\circ}$. As a further more stringent test, we check whether a LC-$\omega$PBE parameterisation of the XC-functional reproduces the QCISD results also for a more intermediate regime. For this purpose, we show charging curves for torsion angles of 110$^{\circ}$ and 130$^{\circ}$ in Fig.~\ref{fig.charge}c, where qualitatively we find that DFT-LC-$\omega$PBE correctly reproduces the gradual change in curve shape from rounded steps to continuously linear with an increase in coupling between the $\pi$ electrons of the phenyl rings, while BHandHLYP does not capture this progression and exhibits straight lines for both angles.

In LC-$\omega$PBE, the exchange component of the Hamiltonian is separated into a short-range (SR) and a long-range (LR) contribution with a range-separation parameter $\omega$ which was empirically fitted to the Gaussian default value of 0.4 Bohr$^{-1}$ (Ref.~\cite{scuseria2}), which can be further refined~\cite{rohrdanz,baer}. This scheme, which uses pure PBE for the SR part and 100\% HF exchange for the LR part, was originally developed to correct the long-range behaviour of the XC-potential for molecules~\cite{scuseria1} but has also been recently found to reduce SIE systematically in systems with fractional electron numbers~\cite{scuseria3}; following the logic of Cohen et al.~\cite{cohen2}, this should also improve the description of the dissociation curve of molecular ions such as H$_2^+$. The explanation for this benign property of the functional offered by the developers of LC-$\omega$PBE is that while a hybrid functional with less than 100\% of HF exchange does not satisfy any universal constraint beyond those satisfied by the underlying semi-local functional, the LR-corrected LC-$\omega$PBE in contrast recovers the exact asymptote of the exchange potential in molecules, which is expected to improve the description of the density tail regions~\cite{scuseria3}. It is exactly this LR asymptotic behavior of the exchange potential which matters for both H$_2^+$ dissociation and the relative stability of fractional charges~\cite{perdew1,perdew}.
Our study thus indicates that long-range corrected hybrid functionals such as LC-$\omega$PBE are best suited for investigating the electron transfer through biphenyl molecules with near perpendicular orientation of the rings. To the best of our knowledge, such functionals although proposed for other purposes have not yet been implemented in any of the common NEGF-DFT codes.
\end{section}

\begin{section}{Summary and discussion}\label{sec:summary}

In summary, we established a connection between intra-molecular electron transfer in biphenyl cations and neutral molecules and the well-studied dissociation problems for H$_2^+$ and H$_2$. It is known for the former that DFT methods with semi-local XC functionals suffer from a delocalization error - a deficiency of the exchange part of the Hamiltonian - while for the latter case, they are inadequately describing static correlation effects. In order to make such a link, we took an intermediate step and considered first schematically the electron transfer inside H$_2$ and H$_2^+$ as well as in the biphenyl molecule and cation with a weak coupling between the components. Thereby, we could make a distinction between the case: (i) of H$_2$ and the neutral biphenyl molecule, where charges have to be created locally in an originally closed shell system and static correlation errors dominate (we have already studied this case extensively in our previous work~\cite{multidet}) and (ii) H$_2^+$ and the biphenyl cation where an already existing charge is just localized on one component by symmetry breaking upon application of the electric field and exchange; for BP$^+$ also dynamical correlation needs to be correctly described in order to avoid the delocalization error and artefactual discontinuities in the charging curves. In order to further investigate the latter scenario, we performed quantum-chemical calculations on a biphenyl cation with various torsion angles and showed that the long-range corrected hybrid functional LC-$\omega$PBE is a good parameterisation for the XC-functional within DFT in this context. 

In this contribution, we investigated the transfer of partial charges from one component of a molecular cation to another, as induced by an external electric field. Although such a study gives an insight into the ability of a theoretical approach to describe the transparency of a coupling barrier inside a molecule which is relevant for electron transport through a single molecule junction, it is clearly not directly comparable to a theoretical description of the transport problem as provided by a NEGF-DFT scheme~\cite{atk}-~\cite{kristian}, where the electrodes are treated explicitly as metal slabs made semi-infinite by Bloch-type conditions. For the situation of a single intra-molecular barrier discussed in our article, however, the similarities between intra-molecular electron transfer and electron transport through a junction in terms of methodological requirements are significant.
In the case of charge transfer inside a biphenyl cation, no local charges are created by the field and therefore no charging energy is involved. Similarly, in a transport setup for a biphenyl molecule bonded to gold electrodes by thiol anchors~\cite{biph1}-~\cite{biph6}, assuming that the thiol groups ensure ideal coupling of the biphenyl MOs to the leads, perpendicular rings would create a situation where two semi-infinite electron densities are separated by one coupling barrier. In such a system there would be no electron localisation on one of its components with a finite extension and therefore no charging energy U, which suggests that an approach with an accurate description of exchange and dynamical correlation such as NEGF-DFT with LC-$\omega$PBE for which we make the case in our article but also the GW method~\cite{thygesen} might be sufficient for a proper treatment of this scenario. Interestingly, we also predict that a method which is good at capturing static correlation such as CASSCF fails to describe dynamical correlation correctly in such a single barrier setup.

As validation for our assumption we refer to Ref.~\cite{biph5}, where the transmission of biphenyl with thiolate and carbodithiolate anchors (BDT and BDCT, respectively) was calculated as a function of the torsion angle with NEGF-DFT. In this work, it was found that the zero-bias conductance with two coplanar phenyl rings differs almost by an order of magnitude with BDCT being more conductive. However, as the torsion angle approaches 90$^{\circ}$, not only the conductance of both molecules in the junction decreases, but also the difference between BDCT and BDT fades out in such a way that the conductance becomes practically the same for the perpendicular conformation. We would interpret this finding as an indication that, as the coupling between the $\pi$-systems of the phenylenes decreases, the limiting barrier for conductance gradually shifts from that determined by the anchors (and favoring BDCT) to that determined intramolecularly by phenylenes (and thus equal for both). This vindicates our assumption above that in a transport setup for such a scenario the anchors to the leads can be regarded as ideal contacts, which makes it in effect a single-barrier system. This observation can serve as a justification for concluding from our molecule-centered approach to charge transfer that in electron transport calculations with NEGF-DFT rounded off steps would be found in the I/V curve for a single molecule junction with a substantial intra-molecular barrier (weak coupling) if an LC-$\omega$PBE (or equivalent) functional is used to ensure a proper exchange description. In order to observe this feature experimentally, a three-electrode scenario with a gate in addition to source and drain would be required (which is the standard for measurements on single electron transistors) for justifying our assumption that the molecule is already charged when the tunnelling process through the single barrier is initiated by the field.
\end{section}


\begin{acknowledgments}
R.S. is currently supported by the Austrian Science Fund FWF, project Nr. P22548, and V.G. by the Interuniversity Attraction Pole IAP 6/27 Program of the Belgian Federal Government. J.C. is funded by the Belgian National Fund for Scientific Research (FNRS).
\end{acknowledgments}


\bibliographystyle{apsrev}

\end{document}